\RequirePackage{ifpdf}
\ifpdf % We are running pdfTeX in pdf mode
\documentclass[pdftex]{sigma}
\else
\documentclass{sigma}
\fi

\begin{document}

\renewcommand{\PaperNumber}{023}

\FirstPageHeading

\ShortArticleName{Characteristic Algebras of Fully Discrete Hyperbolic
Type Equations}

\ArticleName{Characteristic Algebras of Fully Discrete\\ Hyperbolic
Type Equations}

\Author{Ismagil T. HABIBULLIN}
\AuthorNameForHeading{I.T. Habibullin}

\Address{Institute of Mathematics,
 Ufa Scientific Center, Russian Academy of Sciences,\\  112 Chernyshevski
Str., Ufa, 450077 Russia}

\Email{\href{mailto:ihabib@imat.rb.ru}{ihabib@imat.rb.ru}}

\ArticleDates{Received August 04, 2005, in final form November 30,
2005; Published online December 02, 2005}

\Abstract{The notion of the characteristic Lie algebra of the
discrete hyperbolic type equation is introduced. An effective
algorithm to compute the algebra for the equation given is
suggested. Examples and further applications are discussed.}


\Classification{37D99; 37K60; 39A12}

\section{Introduction}
It is well known that the characteristic Lie algebra introduced by
A.B.~Shabat in 1980, plays the crucial role in studying the
hyperbolic type partial differential equations. For example, if the
characteristic algebra of the equation is of finite dimension, then
the equation is solved in quadratures, if the algebra is of finite
growth then the equation is integrated by the inverse scattering
method. More details and references can be found in~\cite{ls}.
Recently it has been observed by A.V.~Zhiber that the characteristic
algebra provides an effective tool for classifying the nonlinear
hyperbolic equations. Years ago the characteristic algebra has been
used to classify integrable systems of a special type~\cite{sy}.
However, the characteristic algebras have not yet been used to
study the discrete equations, despite the fact that the discrete equations had
become very popular the last decade (see, for instance, survey~\cite{zab}).

In this paper we show that the characteristic algebra can be
defined for any discrete equation of the hyperbolic type and it
inherits most of the important properties of its continuous
counterpart. However, it has essentially more complicated
structure. The work was stimulated by~\cite{w}, where the discrete
field theory is studied and the question is posed which types of
algebraic structures are associated with the finite field discrete
3D Toda chains.

\section{Invariants and vector fields}

Consider a discrete nonlinear equation of the form
\begin{gather}\label{eq}
t(u+1,v+1)=f\big(t(u,v),t(u+1,v),t(u,v+1)\big),
\end{gather}
where $t=t(u,v)$ is an unknown function depending on the integers
$u$, $v$, and $f$ is a smooth function of all three arguments. The
following notations are used to shorten formulae: $t_u=t(u+1,v)$,
$t_v=t(u,v+1)$, and $t_{uv}=t(u+1,v+1)$. By using these notations
one can rewrite the equation (\ref{eq}) as follows
$t_{uv}=f(t,t_u,t_v)$. Actually, the equation (\ref{eq}) is a
discrete analog of the partial differential equations.
Particularly, the class of equations (\ref{eq}) contains difference
schemes for the hyperbolic type PDEs on a quadrilateral grid.

The notations above are commonly accepted, but not very convenient
to indicate the iterated shifts. Below we use also different ones.
Introduce the shift operators $D$ and $\bar D$, which act as follows
$Df(u,v)=f(u+1,v)$ and $\bar Df(u,v)=f(u,v+1)$. For the iterated
shifts we introduce the notations $f_j=D^j(f)$ and $\bar f_j=\bar
D^j(f)$, so that $t(u+1,v)=t_1$, $t(u,v+1)=\bar t_1$,
$t(u+2,v)=t_2$, $t(u,v+2)=\bar t_2$ and so on.

The equation (\ref{eq}) is supposed to be hyperbolic. It means
that it can be rewritten in any of the forms:
$t_{u}=g(t,t_{v},t_{uv})$, $t_{v}=r(t,t_{uv},t_u)$, and
$t=s(t_u,t_{uv},t_v)$ with some smooth functions $g$, $r$, and~$s$.

A function $F=F(v,t,t_1,\bar t_1, \dots)$, depending on $v$ and a
finite number of the dynamical variables is called $v$-invariant,
if it is a stationary ``point'' of the shift with respect to $v$ so
that (see also \cite{as})
\begin{gather}\label{int}
\bar DF(v,t,t_1,\bar t_1, \dots)=F(v,t,t_1,\bar t_1, \dots),
\end{gather}
and really the function $F$ solves the equation
$F(v+1,t_v,f,f_1,\bar t_2, \dots)=F(v,t,t_1,\bar t_1, \dots)$.
Exa\-mi\-ning carefully the last equation one can find that:

\begin{lemma}\label{lemma1} The $v$-invariant does not
depend on the variables in
the set $\left\{\bar t_j\right\}_{j=1}^{\infty}$.
\end{lemma}

If any $v$-invariant is found, then each solution of the equation
(\ref{eq}) can be represented as a solution of the following
ordinary discrete equation $F(t,t_u,\dots,t_j)=c(u)$, where $c(u)$ is
an arbitrary function of $u$.

Due to the Lemma~\ref{lemma1} the equation (\ref{int}) can be rewritten as
\begin{gather*}%\label{int1}
F(v+1,t_v,f,f_1, \dots)=F(v,t,t_1,t_2, \dots).
\end{gather*}
The left hand side of the equation contains $t_v$, while the right
hand side does not. Hence the total derivative of $\bar D F$ with
respect to $t_v$ vanishes. In other words, the operator $X_1=\bar
D^{-1}\frac{d}{dt_v} \bar D$ annihilates the $v$-invariant $F$:
$X_1F=0.$ In a similar way one can check that any operator of the
form $X_j=\bar D^{-j}\frac{d}{d t_v} \bar D^{j}$, where $j\geq1$,
satisfies the equation $X_jF=0.$ Really, the right hand side of the
equation $\bar D^jF(v,t,t_1,\bar t_1, \dots)=F(v,t,t_1,\bar t_1,
\dots)$ (which immediately follows from~(\ref{int})) does not depend
on $t_v$ and it implies the equation $X_jF=0$. As a result, one
gets an infinite set of equations for the function $F$. For each
$j$ the operator $X_j$ is a vector field of the form
\begin{gather}\label{vf}
X=\sum^{\infty}_{j=0}x(j)\frac{\partial}{\partial t_j}.
\end{gather}

Consider now the Lie algebra $L_v$ of the vector fields generated
by the operators $X_j$ with the usual commutator of the vector
fields $[X_i,X_j]=X_iX_j-X_jX_i$. We refer to this algebra as
characteristic algebra of the equation (\ref{eq}).

\begin{remark}\label{remark1}
Note that above we started to consider the discrete equation of the
form (\ref{eq}) conjecturing that it admits nontrivial
$v$-invariant. The definition of the algebra was motivated by the invariant.
However the characteristic Lie algebra is still correctly defined
for any equation of the form (\ref{eq}) even if it does not admit
any invariant.
\end{remark}

\section{Algebraic criterion of existence of the invariants}

\begin{theorem}\label{theorem1}
The equation \eqref{eq} admits a nontrivial $v$-invariant if and
only if the algebra $L_v$ is of finite dimension.
\end{theorem}

\begin{proof}
Suppose that the equation (\ref{int}) admits a non-constant
solution. Then the following system of equations
\begin{gather}\label{xeq}
X_jF(v,t,t_1,t_2,\dots,t_N)=0, \qquad j\geq1
\end{gather}
has a non-constant solution. It is possible only if the linear
envelope of the vector-fields $\{X_j\}_{j=1}^{\infty}$ is of
finite dimension.

It is worth mentioning an appropriate property of the vector
fields above. If for a fixed $j$ the operator $X_j$ is linearly
expressed through the operators $X_1, X_2, \dots, X_{j-1}$, then
any operator $X_k$ is a linear combination of these operators.
Really, we are given the expression
$X_j=a_1X_1+\dots+a_{j-1}X_{j-1}$. Note that $X_{j+1}=\bar
D^{-1}X_j\bar D$, hence
$X_{j+1}=D^{-1}(a_1)X_1+\dots+D^{-1}(a_{j-1})X_{j}$. Thus, in this
case the characteristic algebra is generated by the first $j-1$
operators which are linearly independent. Due to the classical
Jacoby theorem the system (\ref{xeq}) has a non-constant solution
only if dimension of the Lie algebra generated by the vector
fields $X_i$ is no greater than~$N$. Thus, one part of the theorem
is proved.

Suppose now that the dimension of the characteristic algebra is
finite and equals, say, $N$, show that in this case the equation
(\ref{eq}) admits a $v$-invariant. Evidently, there exists a
function $G(t,t_1,\dots,t_N)$, which is not a constant and that
$XG=0$ for any $X$ in $L_v$. Such a function is not unique, but
any other solution is expressed as $h(G)$. Due to the construction
the map $X\rightarrow\bar D^{-1}X\bar D$ leaves the algebra
unchanged, hence $G_1=\bar DG$ is also a solution  of the same
system $XG=0$ and therefore $G_1=h(G)$. In other words, one gets a
discrete first order equation: $\bar DG=h(G)$, write its general
solution in the following form: $C=F(v,G)$ where $C$ does not
depend on $v$. Evidently the function $F$ found is just the
$v$-invariant needed.
\end{proof}

\section{Computation of the characteristic algebra}

In this section the explicit forms of the operators $\{X_j\}$ are
given. We show that the operators are vector fields and give a
convenient way to compute the coefficients of the expansion
(\ref{vf}).

Start with the operator $X_1$. Directly by definition one gets
$X_1F(t,t_1,\dots)=\bar D^{-1}\frac{\partial}{\partial
t_v}F(t_v,f$, $f_1,\dots)$. Computing the derivative by the chain rule
one obtains
\begin{gather*}
X_1F(t,t_1,\dots)=\bar D^{-1}\left(\frac{\partial}{\partial
t_v}+\frac{\partial f}{\partial t_v}\frac{\partial}{\partial
f}+\frac{\partial f_1}{\partial t_v}\frac{\partial}{\partial
f_1}+\cdots\right)F(t_v,f,f_1,\dots),
\end{gather*}
 and finally
\begin{gather}\label{1}
X_1=\frac{\partial}{\partial t}+\bar D^{-1}\left(\frac{\partial
f}{\partial t_v}\right)\frac{\partial}{\partial t_u}+ \bar
D^{-1}\left(\frac{\partial f_1}{\partial
t_v}\right)\frac{\partial}{\partial t_2}+ \dots + \bar
D^{-1}\left(\frac{\partial f_j}{\partial
t_v}\right)\frac{\partial}{\partial t_{j+1}}+\cdots .
\end{gather}
So the operator $X_1$ can be represented as
$X_1=\sum\limits_{i=0}^{\infty}x_i\frac{\partial}{\partial t_i}$ where the
coefficients $x_j$ are found as $x_j=\bar
D^{-1}\left(\frac{\partial f_{j-1}}{\partial t_v}\right)$ for $j>0$
and $x_0=1$. Actually the coefficients can be computed by the
following more convenient formula
\begin{gather}\label{xxx}
x_{j+1}=x_1D(x_1)D^2(x_1) \cdots D^j(x_1),
\end{gather}
or $x_{j+1}=x_{j}D^j(x_1)$. Really,
\begin{gather*}
\bar Dx_{j+1}=\frac{\partial
f_j}{\partial t_v}=\frac{\partial f_j}{\partial
f_{j-1}}\frac{\partial f_{j-1}}{\partial f_{j-2}}\cdots\frac{\partial
f}{\partial t_v}=D^j\left(\frac{\partial f}{\partial t_v}\right)\cdots
D\left(\frac{\partial f}{\partial t_v}\right)\frac{\partial f}{\partial
t_v}\\
\phantom{\bar Dx_{j+1}}{}=\bar D\big(D^j(x_1)\cdots D(x_1)x_1\big).
\end{gather*}

To find $X_2$, use the following formula
\begin{gather*}
X_2F=\bar D^{-1}X_1\bar
DF(t,t_1,\dots)=\bar D^{-1}\left(\frac{\partial}{\partial
t}+x_1\frac{\partial}{\partial t_u}+x_2\frac{\partial}{\partial
t_2}+ \cdots\right)F(t_v,f,f_1,\dots).
\end{gather*}
 After opening the parentheses and
some transformation the right hand side of the last formula gets
the form
\begin{gather*}
X_2F(t,t_1,\dots)=\bar D^{-1}\left(X_1(f)\frac{\partial}{\partial f}+
X_1(f_1)\frac{\partial}{\partial
f_1}+X_1(f_2)\frac{\partial}{\partial
f_2}+\cdots\right)F(t_v,f,f_1,\dots)
\end{gather*}
 So the operator can be written
as
\begin{gather}\label{2}
X_2=\bar D^{-1}\left(X_1(f)\right)\frac{\partial}{\partial t_1}+
\bar D^{-1}\left(X_1(f_1)\right)\frac{\partial}{\partial t_2}+
\bar D^{-1}\left(X_1(f_2)\right)\frac{\partial}{\partial
t_3}+\cdots.
\end{gather}
Continuing this way, one gets
\begin{gather}\label{j}
X_{j}=\bar D^{-1}\left(X_{j-1}(f)\right)\frac{\partial}{\partial
t_1}+ \bar D^{-1}\left(X_{j-1}(f_1)\right)\frac{\partial}{\partial
t_2}+ \bar D^{-1}\left(X_{j-1}(f_2)\right)\frac{\partial}{\partial
t_3}+\cdots .
\end{gather}

One can derive an alternative way to compute the coefficients of
the vector fields $X_k$ above. Represent the operators as follows
\begin{gather}\label{coef}
X_k=\sum^{\infty}_{j=0}n_{kj}\frac{\partial}{\partial t_j}.
\end{gather}
We will show that the coefficients $n_{kj}$ of the operators
satisfy the following linear equation
\begin{gather}\label{linh}
\bar Dn_{k+1,j+1}=D^j\left(\frac{\partial f}{\partial
t}\right)n_{k,j}+D^j\left(\frac{\partial f}{\partial t_1}\right)n_{k,j+1}+
D^j\left(\frac{\partial f}{\partial t_v}\right)\bar Dn_{k+1,j},
\end{gather}
closely connected with the direct linearization of the initial
nonlinear equation (\ref{eq}). In order to derive this formula,
apply the operator $X_k$ to the iterated shift
$f_j=f(t_j,t_{j+1},f_{j-1})$ and use the chain rule
\begin{gather}\label{linh2}X_k(f_j)=D^j\!\left(\frac{\partial f}{\partial
t}\right)\!\bar D^{-1}(X_{k-1}(f_{j-1}))+D^j\!\left(\frac{\partial f}{\partial
t_1}\right)\!\bar D^{-1} X_{k-1}(f_j)+ D^j\!\left(\frac{\partial f}{\partial
t_v}\right)\!X_{k}(f_{j-1}).\!\!\!\!
\end{gather}
Comparison of the two representations (\ref{j}) and (\ref{coef}) of
the operator $X_k$ yields $X_{k}(f_{j})\!=\!\bar D n_{k+1,j+1}$. By
replacing in (\ref{linh2}) $X$ with $n$ one gets the formula
(\ref{linh}) required.

The characteristic algebra is invariant under the map
$X\rightarrow D^{-1}XD$. First prove the formula
$D^{-1}X_1D=D^{-1}(x_1)X_1$. To this end use the coordinate
representation of the operator
$X_1=\sum\limits_{i=0}^{\infty}x_i\frac{\partial}{\partial t_i}$ and the
formula (\ref{xxx}) for the coefficients. Then check that
$D^{-1}X_2D=\rho X_2$. Really,
\begin{gather*}
 D^{-1}X_2D=D^{-1}\bar D^{-1}X_1\bar D D=
\bar D^{-1}D^{-1}X_1D\bar D=\bar D^{-1}\big(D^{-1}(x_1)\big)X_2.
\end{gather*}
Obviously similar representations can be derived for any generator
of the characteristic algebra.

The following statement turns out to be very useful for studying
the characteristic algebra.

\begin{lemma}\label{lemma2}
 Suppose that the vector field
\begin{gather*}%\label{4}
X=\sum^{\infty}_{j=0}x_{j}\frac{\partial}{\partial t_j}
\end{gather*}
satisfies the equation
\begin{gather}\label{41}
D(X)=X,
\end{gather}
then $X\equiv0$.
\end{lemma}

\begin{proof}

 It follows from (\ref{41}) that
\begin{gather*}
\sum^{\infty}_{j=0}D\big(x_{j}(\bar t_1, \bar t_{-1},
t,t_1,t_2,\dots,t_{k_j})\big)\frac{\partial}{\partial
t_{j+1}}=\sum^{\infty}_{j=0}x_{j}(\bar t_1, \bar t_{-1},
t,t_1,t_2,\dots,t_{k_j})\frac{\partial}{\partial t_j}.
\end{gather*} Comparison
of the coefficients before the operators $\frac{\partial}{\partial
t_j}$ in both sides of this equation yields: $x_0=0$, $x_1=0$, \dots\
and so on.
\end{proof}

\section{The commutativity property of the algebra}

One of the unexpected properties of the characteristic Lie algebra
is the commutativity of the operators $X_j$. Consider first an
auxiliary statement.

\begin{lemma}\label{lemma3}
The coefficients $x_{ki}$ of the vector fields $X_{k}$, $k\geq 1$,
$i\geq 0$ do not depend on the variable~$t_v$.
\end{lemma}
In other words, the coefficients of the expansions
\begin{gather*}
X_k=\sum^{\infty}_{j=0}x_{kj}\frac{\partial}{\partial t_j}
\end{gather*}
satisfy the equation $\frac{d}{d t_v}x_{ki}=0.$
\begin{proof}
For the operator $X_1$ one has $x_{1,j+1}=\bar
D^{-1}\left(\frac{\partial f_{j}}{\partial t_v}\right)$. But the
function $f_j$ does not depend on $\bar t_2=\bar Dt_v$ and on $\bar
t_3,\bar t_4,\dots$ as well. Hence the coefficients do not depend on
$\bar t_1=t_v$. Similarly, the functions $X_1(f_{j})$ do not depend
on $\bar t_2$, so $x_{2,j+1}=\bar D^{-1}\left(X_1(f_{j})\right)$ do
not contain $t_v$. One can complete the proof by using induction
with respect to $k$.
\end{proof}
Now return to the main statement of the section.
\begin{theorem}\label{theorem2}
For any positive integers $i$, $j$ the equation holds $[X_i,X_j]=0$.
\end{theorem}

\begin{proof} Remind that $X_j=\bar D^{-j}\frac{d}{d t_v} \bar
D^{j}$, so that one can deduce
\begin{gather*}
[X_i,X_j]=\left[\bar D^{-i}\frac{d}{d
t_v} \bar D^{i},\bar D^{-j}\frac{d}{d t_v} \bar D^{j}\right].
\end{gather*}
Suppose
for the definiteness that $k=j-i\geq 1$. Then
\begin{gather*}
[X_i,X_j]=\bar
D^{-i}\left[\frac{d}{d t_v},\bar D^{i-j}\frac{d}{d t_v} \bar D^{j-i}\right]
\bar D^{i}=\bar D^{-i}\left[\frac{d}{d t_v},X_k\right] \bar D^{i}=\bar
D^{-i}\sum\limits^{\infty}_{j=0}\frac{d}{d
t_v}(x_{kj})\frac{\partial}{\partial t_j}\bar D^{i}.
\end{gather*}
 But due to
the Lemma \ref{lemma3} the last expression vanishes that proves the theorem.
\end{proof}

\section{Characteristic algebra for the discrete Liouville equation}

The well known Liouville equation $\frac{\partial^2 v}{\partial
x\partial y}=e^v$ admits a discrete analog of the form \cite{hir}
\begin{gather}\label{L1}
t_{uv}=\frac{1}{t}(t_u-1)(t_v-1),
\end{gather}
which can evidently be rewritten as
\begin{gather*}%\label{L2}
t_{u,-v}=\frac{1}{t-1}t_ut_{-v}+1.
\end{gather*}
Specify the coefficients of the expansions (\ref{1})--(\ref{2})
representing the vector fields $X_{1}$ and $X_{2}$ for the
discrete Liouville equation (\ref{L1}). Find the coefficient
$x_{1}=\bar D^{-1}\left(\frac{\partial f}{\partial t_v}\right)$,
remind that $f(t,t_u,t_v)=\frac{1}{t}(t_u-1)(t_v-1)$ and
$g(t_{-v},t,t_u)=\frac{1}{t-1}t_ut_{-v}+1$,
\begin{gather*}
 x_{1}=\bar D^{-1}\left(\frac{\partial f}{\partial t_v}\right)=
\bar D^{-1}\left(\frac{t_u-1}{t}\right)=\frac{g-1}{t_{-v}}=
\frac{t_u}{t-1}.
\end{gather*}
 Similarly
\begin{gather*}
x_{2}=\bar D^{-1}\left(\frac{\partial f_1}{\partial f}
\frac{\partial f}{\partial
t_v}\right)=\frac{t_2}{t_1-1}\frac{t_1}{t-1}.
\end{gather*}
It can easily be proved by induction that
$x_{j}=\prod\limits_{k=0}^{j-1}\frac{t_{k+1}}{t_k-1}$ for $j\geq1$, here
$t_0:=t$. Remind also that $x_{0}=1$. So that the vector field is
\begin{gather*}%\label{LX1}
X_1=\frac{\partial}{\partial
t}+\frac{t_1}{t-1} \frac{\partial}{\partial t_1}+
\frac{t_1t_2}{(t-1)(t_1-1)} \frac{\partial}{\partial t_2}+
\frac{t_1t_2t_3}{(t-1)(t_1-1)(t_2-1)}
\frac{\partial}{\partial t_3}+\cdots,
\end{gather*}
It is a more difficult problem to find the operator $X_2$, it can
be proved that
\begin{gather*}%\label{LX2}
X_2=\frac{(t-1)t}{(t_{-v}-1)t_{-v}}(X_1-X_{-1}),
\end{gather*}
where the operator $X_{-1}$ is defined as follows $X_{-1}=\bar
D\frac{d}{d\bar t_{-1}} \bar D^{-1}$. For the coefficients
$x_{-,j}$ of the expansion
$X_{-1}=\sum\limits^{\infty}_{j=0}x_{-,j}\frac{\partial}{\partial t_j}$
one can deduce the formula
$x_{-,j}=\prod\limits_{k=0}^{j-1}\frac{t_{k+1}-1}{t_k}$, so that the
operator is represented as
\begin{gather*}%\label{L-1}
X_{-1}=\frac{\partial}{\partial
t}+\frac{t_1-1}{t}\frac{\partial}{\partial t_1}+
\frac{(t_1-1)(t_2-1)}{tt_1}\frac{\partial}{\partial t_2}+
\frac{(t_1-1)(t_2-1)(t_3-1)}{tt_1t_2}
\frac{\partial}{\partial t_3}+\cdots,
\\
X_3=\bar D^{-1}X_2\bar D=\frac{(\bar t_{-1}-1)\bar t_{-1}}
{(\bar t_{-2}-1)\bar t_{-2}} X_2.\nonumber
\end{gather*}

\begin{theorem}\label{theorem3}
Dimension of the characteristic Lie algebra of the Liouville
equation \eqref{L1} equals two.
\end{theorem}

\begin{proof}
It is more easy to deal with the operators $Y_+=(t-1)X_1$ and
$Y_-=tX_{-1}$ rather than the operators $X_{1}$ and $X_{-1}$. In
order to prove the theorem it is enough to check the formula
\begin{gather}\label{Y}
[Y_+,Y_-]-Y_++Y_-=0.
\end{gather}
Denote through $X$ the left hand side of the equation (\ref{Y})
and compute $D(X)$ to apply the Lemma \ref{lemma2}. It is shown
straightforwardly that
\begin{gather*}
D(Y_+)=\frac{t_1-1}{t_1}\left(Y_+-(t-1)\frac{\partial}{\partial
t}\right) \qquad \mbox{and}\qquad
D(Y_-)=\frac{t_1}{t_1-1}\left(Y_--t\frac{\partial}{\partial t}\right).
\end{gather*}
Compute now the shifted operator $D([Y_+,Y_-])=[D(Y_+),D(Y_-)]$
which can be represented as follows
\begin{gather*}
[D(Y_+),D(Y_-)]=\left[\frac{t_1-1}{t_1}\left(Y_+-(t-1)\frac{\partial}{\partial
t}\right),\frac{t_1}{t_1-1}\left(Y_--t\frac{\partial}{\partial
t}\right)\right]=A_1+A_2+A_3+A_4,
\end{gather*}
 where
 \begin{gather*}
A_1=\left[\frac{t_1-1}{t_1}Y_+,\frac{t_1}{t_1-1}Y_-\right],\qquad
A_2=-\left[\frac{(t_1-1)(t-1)}{t_1}\frac{\partial}{\partial
t},\frac{t_1}{t_1-1}Y_-\right],\\
A_3=-\left[\frac{t_1-1}{t_1}Y_+,\frac{tt_1}{t_1-1}\frac{\partial}{\partial
t}\right], \qquad A_4=\left[\frac{(t_1-1)(t-1)}{t_1}\frac{\partial}{\partial
t},\frac{tt_1}{t_1-1}\frac{\partial}{\partial t}\right].
\end{gather*}
 Direct computations give
\begin{gather*}
A_1=[Y_+,Y_-]-\frac{1}{t_1-1}Y_-\frac{1}{t_1Y_+},\qquad
A_2=\frac{1-t-t_1}{t_1}\frac{\partial}{\partial t},\\
A_3=\frac{1-t-t_1}{t_1-1}\frac{\partial}{\partial t},\qquad
A_4=\frac{\partial}{\partial t}.
\end{gather*}
Summarizing all computations above one gets the result
\begin{gather*}
D[Y_+,Y_-]=[Y_+,Y_-]-\frac{1}{t_1-1}Y_--\frac{1}{t_1}Y_+
+\left(1+\frac{t-1}{t_1}+\frac{t}{t_1-1}\right)\frac{\partial}{\partial t},
\end{gather*}
which implies $D([Y_+,Y_-]-Y_++Y_-)=[Y_+,Y_-]-Y_++Y_-$. Now apply
Lemma \ref{lemma2} to the function $X=[Y_+,Y_-]-Y_++Y_-$ to get $X\equiv0$.
\end{proof}

\section{How to find the invariants?}

In this section two examples of the equations with the invariants
are shown. Start with a simple one.

\begin{example} Consider a linear equation of the form
\begin{gather*}%\label{lin}
t_{uv}=t_u+t_v-t+1
\end{gather*}
so that $f(t,t_u,t_v)=t_u+t_v-t+1$ and
$g(t_{-v},t,t_u)=t_u-t+t_{-v}-1$. It is easy to see that the
algebra is of one dimension
\begin{gather*}
X_1=\frac{\partial}{\partial t}+\frac{\partial}{\partial t_u}+
\frac{\partial}{\partial t_2}+ \cdots, \qquad X_2\equiv0.
\end{gather*}

The first integral of the least order for the equation $X_1F=0$ can
be taken as $G_0=t_1-t$. Evidently it solves the equation $\bar
DG_0=G_0+1$. Now, the invariant is to be taken as $I=t_1-t-v$,
because $\bar DI=t_{uv}-t_v-v-1=t_u-t-v=I$. Any other $v$-invariant
of the equation can be represented as $F=F(I,DI,D^2I,\dots,D^kI)$.
\end{example}

\begin{example} Return to the discrete Liouville equation
discussed above.
Find the intersection of the kernels of the operators $X_{1}$ and
$X_{-1}$, hence they constitute the basis of the characteristic
algebra of the Liouville equation (\ref{L1}). To this end first
solve the equation $X_{1}F=0$ which is reduced to the following
infinite system of the ordinary differential equations:
\begin{gather*}
\frac{dt}{1}=\frac{dt_1}{t_1/(t-1)}=
\frac{dt_2}{t_1t_2/(t-1)(t_1-1)}=\cdots .
\end{gather*}
 So that the invariants of
the vector field $X_{1}$ are $I_0=\frac{t_1}{t-1}$, $I_1=D(I_0)$,
$I_2=D(I_1)$, \dots . Change the variables in the vector fields by
setting $\tilde t=t$, $\tilde t_1=I_0$, $\tilde t_2=I_1$, \dots .
Then one gets
\begin{gather*}
X_1=\frac{\partial}{\partial t}\qquad
\mbox{and} \qquad
X_{-1}={\frac{\partial}{\partial t}-\frac{\tilde
t_1+1}{t(t-1)}\frac{\partial}{\partial\tilde t_1}-\frac{\tilde
t_2+1}{\tilde t_1t(t-1)}\frac{\partial}{\partial\tilde t_2}-
\cdots}.
\end{gather*}
 Now solve the equation ${ \frac{d\tilde
t_2}{d\tilde t_1}=\frac{\tilde t_2+1}{\tilde t_1(\tilde t_1+1)}}$
to find the common solution
$F_0=\left(\frac{t_2}{t_1-1}+1\right)\left(\frac{t-1}{t_1}+1\right)$ of both equations
$X_{1}F=0$ and $X_{-1}F=0$. The invariant is
\begin{gather*}
I(t,t_1,t_2)=F_0(t,t_1,t_2)=\left(\frac{t_2}{t_1-1}+1\right)\left(\frac{t-1}{t_1}+1\right)
\end{gather*}
(see also~\cite{as}). Evidently, the basis of common invariants
can be chosen as follows: $F_0, D(F_0)$, $D^2(F_0), \dots$. It can be
easily verified by the direct computation that $F_0$ solves the
equation $\bar DF_0=F_0$.
\end{example}

\begin{corollary} In the case of the Liouville equation each
solution of the system of the equations $X_{1}F=0$, $X_{-1}F=0$
gives the $v$-invariant.
\end{corollary}

\begin{proof}
Each solution of this system can be represented as
$F=F(F_0,D(F_0),D^2(F),\dots)$, hence
\begin{gather*}
\bar DF\big(F_0,D(F_0),D^2(F),\dots\big)=F\big(\bar D(F_0),\bar D(D(F_0)),\bar D(D^2(F)),\dots\big)\\
\qquad{}=F\big(F_0,D(F_0),D^2(F),\dots\big).\tag*{\qed}
\end{gather*}
\renewcommand{\qed}{}
\end{proof}

\section[Discrete potentiated Korteweg-de Vries equation]{Discrete potentiated Korteweg--de Vries equation}

Discrete equation of the form
\begin{gather}
(t_{uv}-t+p+q)(t_{u}-t_v+q-p)=q^2-p^2 \label{dpkdv2}
\end{gather}
is called the discrete potentiated KdV equation (DPKdV). The
constant parameters $p$, $q$ are removed by a combination of the shift
and scaling transformations of the form
$t(u,v)=h\tilde{t}(u,v)-pv-qu$, where $h^2=q^2-p^2$ if $q^2-p^2>0$
and $h^2=-q^2+p^2$ if $q^2-p^2<0$. After this transformation the
equation (\ref{dpkdv2}) takes one of the forms
\begin{gather*}
(t_{uv}-t)(t_{u}-t_v)=\pm1. %\label{dpkdv1}
\end{gather*}
The algebraic properties of the equation do not depend on the
choice of the sign before the unity, we will take the upper sign.
Study the problem how to describe the characteristic Lie algebra
for the equation
\begin{gather}
(t_{uv}-t)(t_{u}-t_v)=1. \label{dpkdv}
\end{gather}
Represent the equation (\ref{dpkdv}) as $t_{uv}=f(t,t_u,t_v)$,
where $f(t,t_u,t_v)=t+\frac{1}{t_{u}-t_v}$. Set $x=\frac{\partial
f}{\partial t_v}=\frac{1}{(t_v-t_u)^2}$, then $x=(f-t)^2$. Below we
use the upper index to denote the shift with respect to $v$ and
lower index to denote the shift with respect to $u$ so that $\bar
D^{-k}D^jx=x^k_j$. Then due to the general rule the operator $X_1$
is written as
\begin{gather*}%\label{1dpkdv}
X_1=\frac{\partial}{\partial t}+x^{1}\frac{\partial}{\partial t_1}+
x^{1}x^{1}_1\frac{\partial}{\partial t_2}+ \cdots +
x^{1}x^{1}_1\cdots x^{1}_{j}\frac{\partial}{\partial t_{j+1}}+\cdots \,.
\end{gather*}

\begin{theorem}\label{theoremk}
Characteristic Lie algebra of the DPKdV equation \eqref{dpkdv} is
of infinite dimension.
\end{theorem}

\begin{proof}
In the case of the DPKdV equation one gets $\frac{\partial
f}{\partial t}=1$, $\frac{\partial f}{\partial t_1}=-x$,
$\frac{\partial f}{\partial t_v}=x$. The formula~(\ref{linh}) is
specified
\begin{gather*}
\bar Dn_{k+1,j+1}=n_{k,j}-x_jn_{k,j+1}+ x_j\bar
Dn_{k+1,j}.
\end{gather*}
 It is more convenient to write it as
\begin{gather*}%\label{link}
n_{k+1,j+1}=\bar D^{-1}n_{k,j}-x_j^1\bar D^{-1}n_{k,j+1}+
x_j^1n_{k+1,j},
\end{gather*}
remind that $D^{-1}x_j=x_j^1$. It is clear now that the
coefficients are polynomials of the finite number of the dynamical
variables in the set $\{x^i_j\}_{i,j=0}^{\infty}$. Order the
variables in this set according to the following rule:
${\rm ord}(x^n_j)>{\rm ord}(x^n_p)$ if $j>p$ and ${\rm ord}(x^m_j)>{\rm ord}(x^n_p)$ if
$m>n$.
\begin{lemma}\label{lemmak}
For any positive $k,j$ the coefficient $n_{k,j}$ for the operator
$X_k$ in the DPKdV case is represented as
\begin{gather*}
n_{k,j}=m_{k,j}x^k_{j-1}+r_{k,j}
\end{gather*}
where $m_{k,j}$ and $r_{k,j}$ are polynomials of the variables with
the orders less than the order of $x^k_{j-1}$, moreover, $m_{k,j}$
is a monomial.
\end{lemma}
Lemma can be easily proved by using the induction method. It allows
to complete the proof of the theorem. The vector fields
$\{X_k\}_{k=1}^{\infty}$ are all linearly independent.
\end{proof}

Due to the theorem the DPKdV equation do not admit any
$v$-invariant. It is not surprising, because the equation can be
integrated by the inverse scattering method or, in other words, it is $S$-in\-tegrable,
it is well known in the case of partial differential equations that
only $C$-integrable equations (Liouville type) admit such kind objects called $x$- and
$y$-integrals.

\section{Conclusion}

The notion of the characteristic algebra for discrete equations is
introduced. It is proved that the equation is Darboux integrable
if and only if its characteristic algebras in both directions are
of finite dimension. The notion can evidently be generalized to
the systems of discrete hyperbolic equations. It would be useful
to compute the algebras for the periodically closed discrete Toda
equation, or for the finite field discrete Toda equations found in
\cite{w} and \cite{h}, corresponding in the continuum limit to the
simple Lie algebras of the classical series $A$ and $C$.

\subsection*{Acknowledgments}
The author thanks A.V.~Zhiber for fruitful discussions. The work has
been supported by the grant RFBR \# 05-01-00775.

\LastPageEnding

\end{document}